# Rapid Prototyping of a Text Mining Application for Cryptocurrency Market Intelligence

Marek Laskowski[1], Henry M. Kim[2]

[1]*Seneca College, Toronto, Ontario, Canada*
marek.laskowski@senecacollege.ca

[2]*York University, Toronto, Ontario, Canada*
hkim@yorku.ca

**Abstract**

*Blockchain represents a technology for establishing a shared, immutable version of the truth between a network of participants that do not trust one another, and therefore has the potential to disrupt any financial or other industries that rely on third-parties to establish trust. Recent trends in computing including: prevalence of Free and Open Source Software (FOSS); easy access to High Performance Computing (HPC i.e. 'The Cloud'); and increasingly advanced analytics capabilities including Natural Language Processing (NLP) and Machine Learning (ML) allow for rapidly prototyping applications for analysis of trends in the emergence of Blockchain technology. A scalable proof-of-concept pipeline for analysis of multiple streams of semi-structured data posted on social media is demonstrated. Preliminary analysis and performance metrics are presented and discussed. Future work is described that will scale the system to cloud-based, real-time, analysis of multiple data streams, with Information Extraction (IE) (ex. sentiment analysis) and Machine Learning capability.*

## 1. Introduction

A cryptocurrency is a system of token exchange between users underpinned and mathematically verifiable by virtue of the same cryptographic principles that underlie encryption on the Internet. Cryptocurrencies are typically implemented as distributed (Peer-to-Peer) systems based on the same Blockchain technologies that are widely argued to have the potential to revolutionize payment, financial, and monetary systems [1].

This paper introduces a multidisciplinary approach, including Computer Systems Engineering, Software Engineering, Natural Language Processing (NLP) to perform analysis of social network data in order to better understand factors underlying the price and other trends in emerging cryptocurrency markets.

In this context, analysis of multiple data streams has been demonstrated [2] [3], however this work is unique in that it combines publicly available social networking website posts with data it scrapes from the "deep web" [4] or portion information on the Internet that cannot readily be accessed using search engines or indexes.

Although currently limited to Cryptocurrency markets, this project lays the groundwork for the real-time data fusion and analysis of multiple data streams relating to Blockchain-based ecosystems in future work.

## 2. Background & Motivation

Presently the only observable instances of Blockchain technology "in the wild" are Cryptocurrencies, and Cryptocurrency-like instruments, but this is expected to change [5]. Therefore, the study of emerging Cryptocurrency "ecosystems" could be a chance to understand how wider adoption of Blockchain-based technologies may unfold in the future.

### 2.1. Blockchain

The value of Blockchain technology is that it provides a mathematically verifiable means of settling exchanges between counterparties that do not trust one another. The problem of reaching consensus between computers in a trust-less network of computers is known as Byzantine Generals Problem [6] in the context of Distributed Systems [7]; and Blockchain represents the first practical solution to this problem [8][9].

Blockchain has the potential to make a huge splash in the Fintech (Financial Technology) landscape. Blockchain can be used to implement a distributed ledger that could be leveraged by financial institutions to settle transactions. Within the next decade, some argue an estimated $20 Billion USD [10] of overhead could be saved yearly, in bank settlements and securities exchanges by switching to a distributed ledger technology.

Blockchain can be used to control ownership of any asset, even real world assets, following arbitrarily complex rules i.e. "smart contracts" or "programmable money" [5][11]. In some cases, the role traditionally occupied by trusted third parties that charge a premium to assume counterparty risk could be challenged, or certainly made more efficient by employing smart-contracts implemented on Blockchain technology. However, the full extent of technological and organizational (legal/regulatory) challenges to implementing such systems in practice remains unclear. To better understand any barriers to adoption, contemporary deployments of Blockchain technology can be studied.

## 2.2. Bitcoin

Bitcoin [9] (https://bitcoin.org/en/), and other cryptocurrencies leverage Blockchain technology to implement a distributed ledger, enabling a network of users to maintain and transfer ownership of Bitcoin tokens that are cryptographically verified and cannot be double-spent by virtue of the protocol.

What's different about Bitcoin compared to previous currency systems is that there is no central authority issuing Bitcoin tokens; they are issued according to a formula embedded within the protocol itself. Instead, the distributed ledger is maintained by an adversarial network of computers running the Bitcoin protocol, rather than a central authority. These networked computers form a Peer-to-Peer (P2P) network of "miners" in the case of Bitcoin, and they are rewarded with newly created Bitcoins as compensation for processing transactions and maintaining the integrity of the ledger.

When a bitcoin transaction occurs (e.g. when a user initiates a purchase by sending bitcoins) the sender broadcasts a request to the Bitcoin network. Miners receive and aggregate these requests into a block and "sign" the block by producing a hash, or cryptographic digest of the transactions in the block as well as the hash or signature of the previous block, forming a chain (hence blockchain).

By design there are many possible "correct" versions of the current block's hash so that miners can compete to be the first to compute a "winning" hash and broadcast it along with the miner's identifying address to its peers. Once the block is verified by the network of peers, the reward of 25 Bitcoins (as of May, 2016) is assigned to the address of that miner. This is the way that all Bitcoins have come into existence. The hashing function is relatively cheap to compute and verify; Miners continually try different combinations of padding at the end of the block (called the "nonce") in order to generate a winning hash. Therefore, the more computing power a miner has the more likely they are to generate the winning hash for that block, although it's effectively stochastic.

The difficulty or probability of generating the winning hash corresponds to the number of leading zeros in the hash, as a hash with four leading zeroes (ex. 0000123…) is more difficult, or improbable, to generate than one with three leading zeroes (ex. 000789…). The difficulty is adjusted by the protocol such that a block is created (a winning hash is generated) approximately every 10 minutes. For example, the winning hash of block #412717 mined on May 1, 2016 at 10:31:58 is: 000000000000000000025486306feab0dce320b9220592568 52f9812103c170720.

In accordance with the principles of Game Theory the collective selfish action of the competing miners presents a computationally significant barrier to any attacker attempting to insert spurious transactions or altering existing transactions. In order to alter or insert one transaction, the attacker would have to change that block, and by virtue of the fact that the hash of the previous block is included in the current block, the attacker would also have to alter every block that comes afterwards, requiring more than half of the computing power of the Bitcoin network at any one time.

The implementation of Bitcoin's Blockchain-based distributed ledger is interesting in several ways. It enables a monetary system in which personal allegiance in the case of trusted third parties is replaced with the mathematical confidence of a distributed, cryptographically verifiable, database of transactions; a public database that anyone can view and contribute to, but nobody can tamper with or destroy. Because there's nobody in control, it's as intractable to police or shut down as the Internet itself, however by the same token there's no central authority for consumer-level-users to appeal to for help. For instance, there are no chargebacks, a boon for sellers, but one of the value propositions of credit cards to consumers. Finally, unlike many monetary systems, Bitcoin is designed to be deflationary in the sense that the supply of Bitcoins is finite; the mining reward will continue to decrease according to a schedule proscribed in the protocol, eventually, until a point at which mining will be supported entirely with transaction fees, and no new Bitcoins will enter circulation.

Bitcoin is one of many possible currencies and networks that can be built on top of Blockchain technology. Indeed, numerous of Cryptocurrencies exist, however the capitalization of the most prolific Cryptocurrency after Bitcoin is orders of magnitude smaller, making Bitcoin the largest Blockchain implementation available for study.

## 3. Materials and Methods

We employed a number of tools to gather and analyze data from multiple sources that we describe in this section.

### 3.1. Data Sources

Although the sources of social media data represent semi-structured data sources, each has their own peculiarities and irregularities that must be dealt with and understood when attempting to extract insights from them. Both data sources are potentially complicated by Internet connectivity issues with the usual causes (ex. service availability). Because of this, additional sources of data, paradoxically, are potentially additional sources of uncertainty, and care must be taken to avoid confusing the unavailability of data with a true drop in messaging volume in response to market conditions.

**3.1.1. Twitter.** The microblogging and social networking site, http://www.twitter.com permits users to share short messages called Tweets that are accessible through Twitter's website, and also through a RESTful web API (https://dev.twitter.com/rest/public) [12]. Hashtags, ex. #bitcoin, are used to identify topics or entities (users are identified with the @ symbol), and users can subscribe to or search hashtags and other users.

In order to handle disconnections gracefully, a Python script was written that requests from the Twitter RESTful web API all tweets that have the hashtag #bitcoin or use the word Bitcoin (case is ignored). Data is returned and stored in JSON (BSON) format. If the connection is lost the script attempts to reconnect, but backs off for an increasingly long period of time if successive reconnection attempts fail. This is to avoid "hammering" the Twitter server with many connection attempts in rapid succession, and is generally considered polite in the sense of protocols.

The analysis of Twitter posts for understanding Bitcoin market conditions has been previously demonstrated [13], although not in combination with a Deep Web data source such as IRC, as described below. Twitter data was collected using the described approach for seven months between June 1, 2015 and December 31, 2015.

**3.1.1. Internet Relay Chat (IRC).** Internet Relay Chat (IRC) [14] is an archaic form of internet messaging, however it's analogous to modern instant messaging applications whereas users transiently join groups or channels organized by topic. Channels are named typically by topic and begin with a hash sign (e.g. #bitcoin) and are in this way similar to Twitter hashtags.

IRC servers or networks of servers are accessible by the public using IRC clients adhering to the protocol, at a particular IP address or URI. Freenode (chat.freenode.net) is one such network purposed for discussion of Open Source projects and as such includes channels for Bitcoin, Open Source Blockchain development, and general discussion related to these topics.

In contrast with Twitter, IRC represents a Deep Web [4] data source, in the sense that it is information that is not readily searchable or indexed by search engines. Most IRC clients by default log network and chat messages in similar but ultimately ad-hoc formats, and in this respect different than Tweets. The Konversation IRC client (http://konversation.kde.org/) was used, because it is conveniently packaged with many linux distributions.

In late May 2015, a list of public bitcoin related channels on freenode were compiled. In the end, the following channels were logged: #bitcoin-assets, #bitcoin-otc, #bitcoin-pricetalk, #bitcoin, and #dogecoin was also included in order to sample the Cryptocurrency space outside of Bitcoin (http://dogecoin.com/). IRC logs were filtered to remove network messages such as: Join, Topic, Quit, Mode, Created, Part, Nick, and Notice, as these were not of interest to this study.

IRC data was collected using the described approach for a period of six months and two-weeks between June 1, 2015 and December 12, 2015; ending after some of the studied IRC channels changed their policy to be invite-only (mode +i), preventing continued data collection efforts.

### 3.2. Natural Language Processing (NLP) Pipeline

As Engineers we espouse the principles of software re-use and modularization. GATE (General Architecture for Text Engineering) [15] is an Open Source framework for rapidly prototyping NLP applications. The GATE framework is designed to be highly modularized, permitting the interchangeability of many plugins and components in order to customize capabilities for the particular needs of the target application. Some features that are available out-of-the-box or easily by loading plugins include Information Extraction [16], Machine Learning [17], scalability to High Performance Computing (HPC) or Cloud environments [18], and tools to manage and organize large volumes of textual data into corpora.

We used the TwitIE (Twitter Information Extraction) [19] pre-packaged application pipeline in order to expedite development. TwitIE includes all the basic components we need to get started such as Gazetteers, Tokenizers, Part-of-Speech Taggers, and Named Entity Transducers. The GATE Developer tool is provided with GATE to enable rapid application development by permitting the user to reconfigure existing pipeline components using a graphical user interface (GUI), as well as, permitting the browsing of documents and their annotations interactively. A screen capture of GATE developer is show in Figure 1, illustrating some of the features of this tool. The pipeline can be altered by adding, removing, or reconfiguring components by selecting them from the list of components on the left. On the right hand side we see some example tweets that have been annotated using the pipeline.

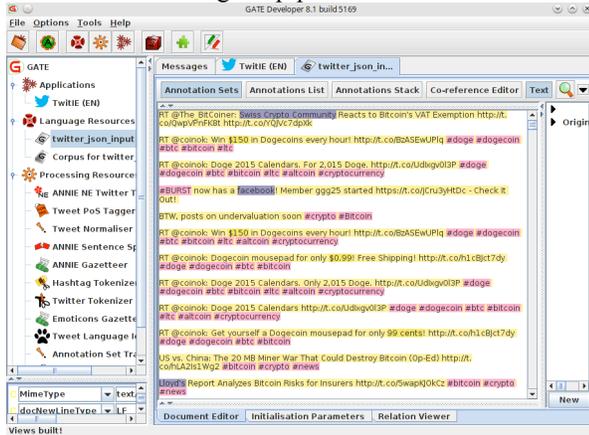

**Figure 1. Gate Developer GUI displaying TwitIE pipeline components (left) and annotated tweets (right).**

Applications developed using GATE Developer can be exported, and the resulting processing pipeline can re-used within any application written in Java (thanks to GATE's Java API) permitting the rapid development of post-processing resources. One such approach treats the annotated document as a graph-like or network structure, which can be traversed and the arbitrarily complex relationships between document (tagged) entities can be examined programmatically.

Before the collected could be processed using the TwitIE pipeline, we discovered we had to perform pre-cleaning and normalization. For example, escaped Unicode symbols such as "\u2026" (representing a horizontal ellipsis) commonly used on the Web had to be cleaned from the data (Ex. replaced by spaces to avoid corrupting the payload size of the tweet data) to prevent the TwitIE pipeline from crashing. A unix pipeline filter program was written for maximum flexibility and later reuse, that reads in an unsanitized JSON file from standard input, and writes the cleaned file to standard output.

The modular architecture of GATE and the developer-centric API, make it possible to rapidly prototype plugins and create processing pipelines for new semi-structured data formats such as IRC logs.

## 4. Results

Our results were ultimately bounded by available compute time. Working on a modest single 2GHz processor core, 71 days of compute time were needed to process the 200GB of data that were collected during the 7 months. GATE used 1.95GB on average from the available 4GB of RAM. As GATE requires a considerable amount of heap memory a 64-Bit Operating System is recommended to permit increased limits on the Java Virtual Machine (JVM) virtual heap space. However, some Operating Systems enforce arbitrary limits to JVM heap space, making memory intensive Java processes such as GATE pipelines somewhat constrained by the operating environment.

**Table 1. Summary of findings; Count of total observed messages, and Pearson product-moment correlation coefficient between messages-per-day in several data streams and Bitcoin market metrics.**

| Data Source | Total Messages | Bitcoin Volume Correlation | Bitcoin Price Correlation |
|---|---|---|---|
| Twitter | 12105833 | 0.5239 | -0.0191 |
| #bitcoin-assets | 189393 | -0.1201 | -0.2991 |
| #bitcoin-otc | 111499 | -0.0568 | -0.0675 |
| #bitcoin-pricetalk | 64712 | 0.7714 | 0.5715 |
| #bitcoin | 214283 | 0.0130 | -0.1355 |
| #dogecoin | 1113243 | -0.1682 | -0.3333 |

Table 1 compares total messages and correlations between daily message counts in each social media stream and Bitcoin market activity indicators. One thing that can be immediately noted from Table 1 is the average volume of Twitter messages is considerably higher than IRC messages. Daily Bitcoin price and trading volume on USD exchanges was downloaded from http://blockchain.info

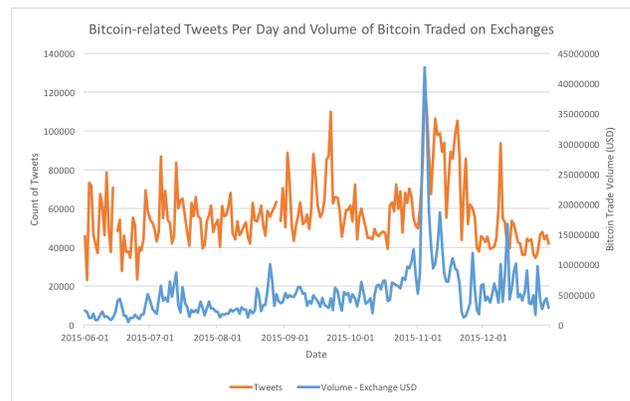

**Figure 2. Count of tweets per day and Bitcoin trading volume as measured by USD exchanges (Source: Blockchain.info).**

Despite efforts to maintain a connection, unexplained significant gaps exist in collected Twitter data June 14, 2015, and August 29, 2015, and in some IRC data between November 19-21, 2016.

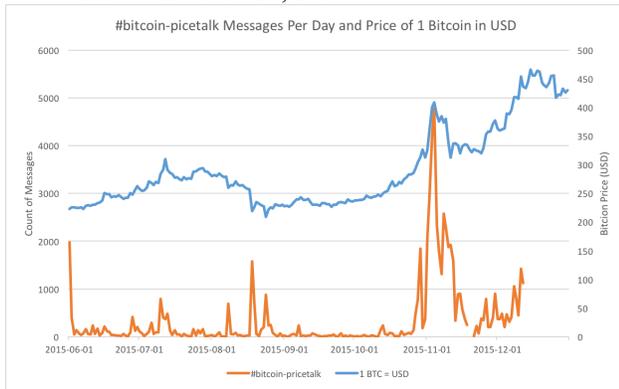

**Figure 3. Count of messages per day in #bitcoin-pricetalk and price of Bitcoin in USD (Source: Blockchain.info).**

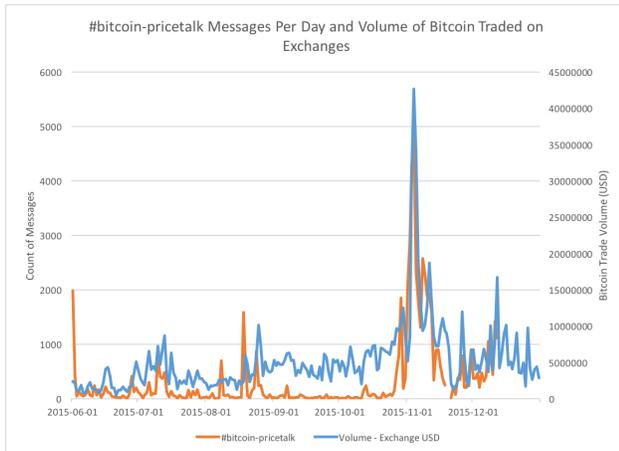

**Figure 4. Count of messages per day in #bitcoin-pricetalk and Bitcoin trading volume as measured by USD exchanges (Source: Blockchain.info).**

## 5. Discussion

According to Table 1 and Figure 2 Twitter message volume is somewhat correlated with the volume of bitcoin transactions on USD-based exchanges as has been noted by previous work [13]. The count of messages per day in the IRC channel #bitcoin-pricetalk, interestingly is positively correlated with both Bitcoin price and trading volume according to Table 1 and Figures 3 and 4 respectively. As of December 2015 #bitcoin-pricetalk has changed its policy to allow only invited users to join.

Some other results bear brief mention; the number of messages per day related to another Cryptocurrency, dogecoin, shows negative correlation with the price of Bitcoin. The number of messages tends to decrease as the price of Bitcoin increases but it remains unclear whether is it due to general trend in dogecoin, or perhaps interest in dogecoin wanes as Bitcoin price increases? We also note there's some negative correlation between bitcoin price and discussion messages per day in #bitcoin-assets. Again it remains unclear whether discussion there has waned because members are satisfied with their Bitcoin assets.

Overall, the results suggest that even with relatively simple measures such as message frequency, we can make some interesting observations. More sophisticated analysis is expected to yield further insights. Furthermore, counterintuitively, despite the volume of messages being lower, owing to the specificity of user interest, and due to the low level of spam, IRC turned out to be quite valuable and a complementary counterpart to Twitter. Again, counterintuitively the constant heartbeat of spam messages on Twitter can be used to infer problems with network performance. Finally, a surprising amount of compute time was required, and the scalability concerns brought to light are to be addressed in Future Work.

## 6. Conclusions & Future Work

We have demonstrated a framework that's capable of fusing multiple semi-structured social media data streams into a coherent picture of a Cryptocurrency marketplace. Some preliminary results demonstrate the utility of this data in understanding complex Blockchain ecosystems.

This lays the groundwork for real-time analysis of multiple data streams, however the computational limits we encountered suggest that we will have to harness additional compute resources within HPC systems, or the Cloud. The GATE framework has been demonstrated as a viable tool for perusing these goals.

Our intention is to extend this work to cover other cryptocurrencies. Using the data we have collected, we will identify additional relevant keywords or hashtags to include as part of data collection. Additional data sources will also be considered such as search engine results and message board postings. Finally, to delve deeper into the semantic meaning of the data, we will perform Information Extraction [20] including but not limited to Sentiment Analysis [21] which may require the creation of domain specific ontologies [22] that would capture the semantic meaning and relationships between entities and concepts in Bitcoin and Blockchain ecosystems.